\documentclass[12pt]{article}
\usepackage{hyperref}
\usepackage{float}
\usepackage{amssymb,amsmath,graphicx,latexsym,cite}

\begin{document}

\title{Vaidya Spacetime for Galileon Gravity's Rainbow}

\author{Prabir Rudra$^{1}$\thanks{prudra.math@gmail.com}~, Mir Faizal$^{2}$ \thanks{mirfaizalmir@gmail.com} and
Ahmed Farag Ali$^{3}$\thanks{ahmed.ali@fsc.bu.edu.eg}\\ $^{1}$\emph{\small{Department of Mathematics, Asutosh College, Kolkata-700 026, India.}} \\
$^{2}$\emph{\small{Department of Physics and Astronomy,
University of Lethbridge, Lethbridge, Alberta, T1K 3M4, Canada}}\\
$^{3}$\emph{\small{Department of Physics, Faculty of Science, Benha University, Benha, 13518, Egypt.}}}

\maketitle

\begin{abstract}
 In this paper, we  analyze Vaidya spacetime with an energy dependent metric in  Galileon gravity's rainbow.
 This will be done using the rainbow
  functions which are  motivated from the results obtained in
  loop quantum gravity approach and   noncommutative geometry.
  We will   investigate the Gravitational collapse in this Galileon gravity's rainbow.
  We will discuss the behavior of   singularities  formed from  the gravitational
  collapse in this rainbow deformed Galileon gravity.
\end{abstract}

\section{Introduction}

The  observations from   type I supernovae indicate that our universe has a positive cosmological constant and
is accelerating in its expansion  \cite{super}-\cite{super5}.
Furthermore, it is known that general theory of relativity has not tested at very large or very small scales, and it is possible
for the general theory of relativity to be modified at such scales.
However, as gravity has been thoroughly tested at the scale of solar system, it is important for
any theory of modified gravity to reduce to    the  general theory of relativity
at the scale of the solar system.
It may be noted an interesting model of modified gravity is called the DGP brane model has been proposed
to explain accelerating cosmic expansion  \cite{Deffayet3}.
This model has two branches and one these branches admits a self-accelerating solution.
However, this model   contains ghost instabilities, and thus cannot be used  as a physical model
for the cosmic acceleration
\cite{Koyama1}.
It may be noted that  such   instabilities also occur for other models of modified gravity \cite{Koyama2}.
Such instabilities occur due to the introduction of extra degrees of freedom into the theory because
of the existence of higher derivative terms.

However, it is possible to construct an
  infrared modification of general theory of relativity \cite{Nicolis1}. This theory
contains a self-interaction term of the form
$\left(\nabla \phi\right)^{2}~^{\fbox{}}~\phi$, and so   general relativity is recovered
 at high densities. It is interesting to note that in the Minkowski background, this theory is   invariant under the
Galileon shift symmetry, $\delta_{\mu}\phi\rightarrow
\delta_{\mu}\phi+c_{\mu}$. This symmetry prevents the occurrence of higher derivative terms in the equation of motion
of this theory. As this theory does not contain extra degrees of freedom,  it cannot also contain
ghost instabilities. The coupling between a Galileon scalar field and massive gravity through composite metrics has also been studied
\cite{k1}.A full set of equations of motion for a flat Friedmann-Robertson-Walker background were obtained
in this theory.  The cosmology has also been studied using  Galileon gravity, and this has been done
by analyzing the   linear perturbation in Galileon gravity \cite{k7}.
Furthermore, low density stars with slow rotation and static relativistic stars have also been analyzed
using Galileon gravity \cite{k2}. It was observed that the the scalar field solution ceases to
exist above a critical density, and this  corresponds to the maximum mass of a neutron star.
The  spherical collapse has also been analyzed in the  Galileon gravity \cite{v2}-\cite{k4}.
This was done by analyzing the solutions to the Einstein equations in Galileon gravity.
Then these solution were used for discussing the
conditions for the formation of a black hole or a naked singularity in Galileon gravity.
In this paper, we shall perform such an analysis in a theory which combines Galileon gravity with
gravity's rainbow.

Another interesting modification to general relativity is called the Horava-Lifshitz gravity \cite{HoravaPRD}-\cite{HoravaPRL}.
This theory of gravity is obtained from a  UV completion of general relativity,  such that general relativity is recovered
in the IR limit
\cite{HoravaPRD}-\cite{HoravaPRL}.  This is done by  taking different Lifshitz scaling for space and time.
Such a different Lifshitz scaling for space and time has also been taken in
type IIA string theory \cite{A}, type  IIB string theory \cite{B},   AdS/CFT correspondence
\cite{ho}-\cite{oh},  dilaton black branes
\cite{d}-\cite{d1}, and dilaton black holes \cite{dh}-\cite{hd}.
The  Horava-Lifshitz gravity is based on the modification of the usual
 usual energy-momentum
dispersion relation in the UV limit such that it reduces to the
usual energy-momentum dispersion relation in the IR limit. The gravity's rainbow is another modification of gravity based on
such a modified energy-momentum
dispersion relation in the UV limit  \cite{MagueijoCQG}-\cite{Amelino2}. In gravity's rainbow
  the metric depends on the energy of the test particle used to probe the structure of the spacetime.
  The  gravity's rainbow can be  related to  the Horava-Lifshitz gravity, for a specific choice of rainbow functions
\cite{re}. There is a strong motivation to study such theories based on the energy-momentum dispersion relation in the UV limit.
This is because the
Lorentz symmetry fixes the form of the energy-momentum relations,
and there are strong theoretical indications
from  various different approaches to quantum gravity that Lorentz symmetry
might only be a symmetry of the low energy effective  field
theory, and so it will break in the UV
limit \cite{1}-\cite{5}. This is expected to occur in  discrete
spacetime \cite{6}, models based on string field theory \cite{7},
spacetime foam \cite{8}, the spin-network in loop quantum gravity
(LQG)\cite{9}, and  non-commutative geometry \cite{10}. It may be noted
that such a  deformation of the standard energy-momentum
dispersion relation in the UV  limit of the theory  leads  the
existence of a maximum energy scale. The doubly special relativity
is build on the existence of such a maximum energy scale
\cite{13}, and gravity's rainbow is the  generalization
of doubly special relativity to curved spacetime  \cite{14}. In
 gravity's rainbow,
the metric describing the geometry  of spacetime   depend on the energy of the test
particle used to probe the structure of that spacetime. So,  the geometry of spacetime is
represented by a family of energy dependent metrics forming a
rainbow of metrics. In gravity's rainbow, the    energy-momentum
dispersion relation is modified by energy dependent rainbow
functions, $F(E)$ and $G(E)$, such that
\begin{equation}  \label{MDR}
E^2F^2(E)-p^2G^2(E)=m^2.
\end{equation}
As it is required that the usual energy-momentum dispersion relation is recovered in the IR limit,
these   rainbow functions    are required  to satisfy
\begin{equation}
\lim\limits_{E/E_P\to0} F(E)=1,\qquad \lim\limits_{E/E_P\to0} G(E)=1.
\end{equation}
The energy dependent metric in gravity's rainbow can be written as
\begin{equation}  \label{rainmetric}
g^{\mu\nu}(E)=\eta^{ab}e^\mu_a(E) e^\nu_b(E).
\end{equation}
The rainbow functions are defined using
the energy $E$, which is the energy at which the spacetime is probed, and  this
energy cannot exceed the Planck energy $E_p$.

Vaidya spacetime  is a  a non-stationary Schwarzschild
spacetime \cite{v1}-\cite{v3}. The gravitational collapse in  Vaidya spacetime has been studied in  Galileon gravity
\cite{v2}. In this paper, we will analyze the gravitational collapse in  Vaidya spacetime in Galileon gravity deformed by rainbow functions. The gravitational collapse has also been studied in   gravity's rainbow \cite{1a}-\cite{2a}.
In fact, the  thermodynamics of black holes has also been discussed in  gravity's rainbow   \cite{16}-\cite{4A}. This has been done by deforming the black hole metric by rainbow functions. The energy $E$ used to define the rainbow functions can be identified with the energy of  quantum particle in the vicinity of the event horizon, which could be emitted in the Hawking radiation. It is possible to obtain a bound  on this energy $E \geq
1/\Delta x $, using the uncertainty principle $\Delta p \geq
1/\Delta x $. Here  the   uncertainty in position of a
particle in the vicinity of the event horizon can  be equated
with  the radius of the event horizon radius
\begin{equation}
 E\geq 1/{\Delta x} \approx 1/{r_+}.
\end{equation}
The existence of this  bound on the energy   modifies the
temperature of the black hole in gravity's rainbow. This    modified temperature of the
black hole  has been used for   calculate the corrected entropy of a
black hole in gravity's rainbow.  This  deformation of the    black
hole thermodynamics lead to the formation of black remnants,   and these black remnants
 can have important phenomenological implication
for the detection of mini black holes at the LHC \cite{g1}.
  It may be noted
that this energy which is used in constructing rainbow functions  dynamically  dependent on the
  coordinate \cite{re}. Even though we do not need this explicit dependence
of this energy  on  the coordinate, but it is important to note that the rainbow functions are dynamical, and so
they cannot be gauged away.

\section{Field Equations and the Solutions in Vaidya Space-time in the
background of Galileon Gravity}\label{calculation}

The Galileon theory is     invariant under the
Galileon shift symmetry. Now if ${\cal L}_{m}$~~ is the matter Lagrangian and $\phi$ is the Galileon field, then
the action for such a theory can be written as   \cite{Nicolis1}-\cite{Silva1},
\begin{equation}\label{Lag}
S=\int d^{4} x \sqrt{-g}\left[\phi R- \frac{w}{\phi} \left(\nabla
\phi\right)^{2}+ f(\phi)^{\fbox{}}~\phi \left(\nabla
\phi\right)^{2}+{\cal L}_{m}\right],
\end{equation}
where  $w$ is the
Galileon parameter, and the    coupling $f(\phi)$ has
dimension of length.  Furthermore, we also have   $\left(\nabla
\phi\right)^{2}=g^{\mu\nu}\nabla_{\mu}\phi \nabla_{\nu}\phi$, and 
$^{\fbox{}}~\phi=g^{\mu\nu}\nabla_{\mu}\nabla_{\nu}\phi$.  
  Now  for a    spherically
symmetric spacetime, we can write    \cite{v1},
\begin{equation}\label{collapse2.1}
ds^{2}=-\left(1-\frac{m(t,r)}{r}\right)dt^{2}+2dtdr+r^{2}d\Omega_{2}^{2}.
\end{equation}
Here   the radial coordinate is denoted by $r$ and   the null
coordinate is denoted by $t$. The  gravitational mass inside the
sphere of radius $r$ is denoted by $m(t,~r)$, and the     line element on
a unit $2$-sphere is denoted by $d\Omega_{2}^{2}$.

The   Rainbow  deformations the above metric can be written as
\begin{equation}\label{rainbow}
ds^2 =-\frac{1}{F^2(E)}\left(1-\frac{m(t,
r)}{r}\right)dt^{2}+\frac{1}{F(E)G(E)}dtdr+\frac{1}{G^2(E)}r^2
d\Omega_2^2.
\end{equation}
The   Einstein's
equations for this metric can be written as
$$G_{\mu
\nu}=\frac{T_{\mu\nu}}{2\phi}+\frac{1}{\phi}\left(\nabla_{\mu}\nabla_{\nu}\phi-g_{\mu\nu}~^{\fbox{}}~\phi\right)
+\frac{\omega}{\phi^{2}}\left[\nabla_{\mu}\phi\nabla_{\nu}\phi-\frac{1}{2}g_{\mu\nu}\left(\nabla\phi\right)^{2}\right]$$
\begin{equation}
-\frac{1}{\phi}\left\{\frac{1}{2}g_{\mu\nu}\nabla_{\lambda}[f(\phi)\left(\nabla
\phi\right)^{2}]\nabla^{\lambda}\phi-\nabla_{\mu}[f(\phi)\left(\nabla
\phi\right)^{2}]\nabla_{\nu}\phi+f(\phi)\nabla_{\mu}\phi\nabla_{\nu}\phi~^{\fbox{}}~\phi\right\}
\end{equation}
 {where $T_{\mu\nu}$ is the energy momentum tensor}.

 The energy-momentum tensor for the Vaidya null radiation is given by
\begin{equation}\label{collapse2.4}
T_{\mu\nu}^{(n)}=\sigma l_{\mu}l_{\nu},
\end{equation}
 where $\sigma$ is the energy density corresponding to
Vaidya null radiation.
 The energy-momentum tensor for
a perfect fluid  is given by
\begin{equation}\label{collapse2.5}
T_{\mu\nu}^{(m)}=(\rho+p)(l_{\mu}\eta_{\nu}+l_{\nu}\eta_{\mu})+pg_{\mu\nu},
\end{equation}
where $\rho$ and $p$ are the energy density  and pressure for the
perfect fluid. Now we can write \cite{Rudra1}
\begin{equation}\label{collapse2.3}
T_{\mu\nu}=T_{\mu\nu}^{(n)}+T_{\mu\nu}^{(m)}
\end{equation}
  It may be noted that    $l_{\mu}$ and $\eta_{\mu}$ are linearly independent
future pointing null vectors,
\begin{equation}\label{collapse2.6}
l_{\mu}=(1,0,0,0)~~~~ and~~~~
\eta_{\mu}=\left(\frac{1}{2}\left(1-\frac{m}{r}\right),-1,0,0
\right).
\end{equation}
Furthermore, they  satisfy
\begin{equation}\label{collapse2.7}
l_{\lambda}l^{\lambda}=\eta_{\lambda}\eta^{\lambda}=0,~
l_{\lambda}\eta^{\lambda}=-1
\end{equation}

Now we can write the  Einstein field equations ($G_{\mu\nu}=T_{\mu\nu}$) for the
metric (\ref{rainbow}),  and the wave equation for the Galileon
field $\phi$. Thus, we can use  $G_{00}=T_{00}$, to obtain
\begin{eqnarray}\label{1}
&& \frac{G(E)\left[G(E)m\left\{3-4m'\right\}+r\left\{-3G(E)+4G(E)m'+2F(E)\dot{m}\right\}\right]}{F(E)^{2}r^{3}}=
\frac{\sigma+\rho\left(1-\frac{m}{r}\right)}
{2\phi} \nonumber \\ &&+\frac{1}{\phi}\Big[\ddot{\phi}-\left(\frac{m}{2r^{2}}-\frac{m'}{2r}\right)\dot{\phi}
 -\left(\frac{m}{2r^{2}}-\frac{m^{2}}{2r^{3}}-\frac{m'}{2r}+\frac{mm'}{2r^{2}} +\frac{\dot{m}}{2r}\right)\phi' \nonumber \\ &&+\left(1-\frac{m}{r}\right)\left\{2\dot{\phi}'-\phi'\left(\frac{m'}{r}
-\frac{3m}{r^{2}}+\frac{2}{r}\right)+\left(1-\frac{m}{r}\right)\phi''\right\}\Big]
 \nonumber \\ && +\frac{\omega}{\phi^{2}}\left[\dot{\phi}^{2}+\frac{1}{2}\left(1-\frac{m}{r}\right)
\phi'\left(2\dot{\phi}+\left(1-\frac{m}{r}\right)\phi'\right)\right]\nonumber \\ &&+\frac{1}{\phi}\Big[\frac{1}{2}
\left(1-\frac{m}{r}\right)\left\{\phi'\nabla_{0}U
+\left(\dot{\phi}+\left(1-\frac{m}{r}\right)\phi'\right)\nabla_{1}U\right\} \nonumber \\ && -\dot{\phi}\nabla_{0}U+f(\phi)\dot{\phi}^{2}\left\{2\dot{\phi}'-\phi'\left(\frac{m'}{r}
-\frac{3m}{r^{2}}+\frac{2}{r}\right)\left(1-\frac{m}{r}\right)\phi''\right\}\Big]
\end{eqnarray}
\vspace{3mm}
We can  use $G_{11}=T_{11}$, to obtain
\begin{eqnarray}\label{2}
\frac{\phi''}{\phi}&+&\frac{\omega\phi'^{2}}{\phi^{2}}-\frac{1}{\phi}\Big[-f(\phi)
\left\{2\phi'^{2}\dot{\phi}'+2\dot{\phi}\phi'\phi''+2\left(1-\frac{m}{r}\right)
\phi'^{2}\phi''+\frac{m}{r^{2}}\phi'^{3}\right\}\nonumber \\&-&f'(\phi)\left\{2\phi'^{3}\dot{\phi}
+\left(1-\frac{m}{r}\right)\phi'^{4}\right\}\nonumber \\&+&f(\phi)\phi'^{2}\left\{2\dot{\phi}'-\phi'\left(\frac{m'}{r}-\frac{3m}{r^{2}}
+\frac{2}{r}\right)+\left(1-\frac{m}{r}\right)\phi''\right\}\Big]=0,
\end{eqnarray}
\vspace{3mm}
We can use $G_{01}=T_{01}$, to  obtain
\begin{eqnarray}\label{3}
&&\frac{G(E)\left\{4m'-3\right\}}{2r^{2}F(E)} \nonumber \\ &&=\frac{\rho}{2\phi}+\frac{1}{\phi}\left[\dot{\phi}'+\phi'\left(\frac{m'}{2r}
-\frac{m}{2r^{2}}\right)-\phi'\left(\frac{m'}{r}-\frac{3m}{r^{2}}+\frac{2}{r}\right)
+\phi''\left(1-\frac{m}{r}\right)\right]\nonumber \\ &&+\frac{\omega}{2\phi^{2}}\left(1-\frac{m}{r}\right)\phi'^{2}
+\frac{1}{\phi}\Big[\frac{1}{2}\nabla_{0}U\left(\dot{\phi}-2\phi'\right)+\frac{1}{2}\phi'\nabla_{1}U
\nonumber \\ &&+f(\phi)\dot{\phi}\phi'\left\{2\dot{\phi}'-\phi'\left(\frac{m'}{r}-\frac{3m}{r^{2}}+\frac{2}{r}\right)
+\left(1-\frac{m}{r}\right)\phi''\right\}\Big]~,
\end{eqnarray}
We can use $G_{22}=T_{22}$, to obtain
\begin{eqnarray}\label{4}
2rm''&=&\frac{\omega}{\phi^{2}}\left[\frac{r^{2}}{2}\phi'\left\{2\dot{\phi}
+\left(1-\frac{m}{r}\right)\phi'\right\}\right]\nonumber \\&-&\frac{1}{\phi}\left[r^{2}\left\{\phi'\left(\frac{m'}{r}
-\frac{3m}{r^{2}}+\frac{2}{r}\right)-\left(1-\frac{m}{r}\right)\phi''-2\dot{\phi}'\right\}\right] \nonumber \\ &&
+\frac{r^{2}}{2\phi}\left(\nabla_{0}U\dot{\phi}+\nabla_{1}U\phi'\right)-\frac{p
r^{2}}{2\phi}.
\end{eqnarray}
Finally, we can use $G_{33}=T_{33}$, to obtain
\begin{eqnarray}\label{5}
\frac{pr^{2}}{2\phi}&+&\frac{1}{\phi}\left[r\dot{\phi}-\left(m-r\right)\phi'-r^{2}\left\{2\dot{\phi}'
\phi'\left(\frac{m'}{r}-\frac{3m}{r^{2}}+\frac{2}{r}\right)+\left(1-\frac{m}{r}\right)\phi''\right\}\right] \nonumber \\ &&
-\frac{\omega}{\phi^{2}}\left[\frac{\phi'}{2}r^{2}\left(2\dot{\phi}+\phi'\left(1-\frac{m}{r}\right)\right)\right]
-\frac{1}{2\phi}r^{2}\left(\dot{\phi}\nabla_{0}U+\phi'\nabla_{1}U\right)+2rm''=0.
\end{eqnarray}
Here the differentiation with respect
to $t$ is denoted by a over-dot and   differentiation with respect $r$ is denoted by a dash. It is useful to define  $U=f(\phi)\left(\nabla
\phi\right)^{2}$. Now we can  write an   expression for
$\nabla_{0}U $ as
\begin{equation}
\nabla_{0}U=f(\phi)\left[2\phi'\ddot{\phi}+2\dot{\phi}\dot{\phi}'
+\left(1-\frac{m}{r}\right)2\phi'\dot{\phi}'-\phi'^{2}\frac{\dot{m}}{r}\right]
+f'(\phi)\left[2\phi'\dot{\phi}^{2}+\left(1-\frac{m}{r}\right)\phi'^{2}\dot{\phi}\right]
\end{equation}
and an expression for $\nabla_{1}U$ as
\begin{equation}
\nabla_{1}U=f(\phi)\left[2\phi'\dot{\phi}'+2\dot{\phi}\phi''+2\left(1-\frac{m}{r}\right)\phi'\phi''+\frac{m}{r^{2}}
\phi'^{2}\right]+f'(\phi)\left[2\phi'^{2}\dot{\phi}+\left(1-\frac{m}{r}\right)\phi'^{3}\right]
\end{equation}
It is difficult to solve these equations explicitly, and so we
  assume $P(r)$ is an arbitrary function of $r$ and $Q(t)$ is an
arbitrary function of $t$, and write
\begin{equation}\label{7}
\phi(r,t)=P(r)Q(t)
\end{equation}
 It may be noted that     $f(\phi)$ is an arbitrary
function of $\phi$, so we can
write,
\begin{equation}
f(\phi)=f_{0}\phi^{-2}
\end{equation}
where, $f_{0}$ is a constant. This is a particular form of Galileon gravity rather than the most general form
of Galileon gravity. As the general form of Galileon gravity was very complicated, we simplified our analysis
by assuming this   particular form of Galileon gravity. It is possible to obtain analytic solutions in this
  particular form of Galileon gravity. We assume that the barotropic equation of state holds for the  matter fluid
\begin{equation}\label{8}
p=k\rho
\end{equation}
where '$k$' is a constant. The solution for $Q(t)$ can be written as
\begin{equation}
Q(t)=\alpha_{1}e^{-\lambda t}
\end{equation}
Here $\alpha_{1}$ and $\lambda$ are arbitrary constants. It is not possible to obtain a similar solution for
   $P(r)$ as the field equations are very
 complicated. So, we assume that
\begin{equation}
P(r)=\alpha r^{n}
\end{equation}
where $\alpha$ and $n$ are arbitrary constants. We use these values of
  $P$ and $Q$ in the field equations {and
considering $f_{0}=1$ (without much loss of generality in the
given context)}. Thus, we obtain the  following  differential equation
$$r^{2}m''+\left[4k\frac{G(E)}{F(E)}+n\left(2+k\right)\right]rm'+\left[n\left\{2\left(k+1\right)\left(n-1\right)
-\left(5k+6\right)\right\}\right]m+2n\left[\left(3-n\right)\left(k+1\right)r\right.$$
\begin{equation}
\left.+\left(\omega+k+2\right)\lambda
r^{2}\right]-3k\frac{G(E)}{F(E)}r=0
\end{equation}
Now we obtain an explicit expression for $m$ by
solving these  differential equations,
\begin{equation}\label{10}
m(t,r)=f_{1}(t)r^{\omega_{1}}+f_{2}(t)r^{\omega_{2}}+\frac{2n\left(n-3\right)\left(k+1\right)+3k\frac{G(E)}{F(E)}}
{\left(1-\omega_{1}\right)
\left(1-\omega_{2}\right)}r-\frac{2n\lambda\left(\omega+k+2\right)}{\left(2-\omega_{1}\right)
\left(2-\omega_{2}\right)}r^{2}
\end{equation}
where
\begin{eqnarray}\label{11}
\omega_{1},
\omega_{2}&=& \left[1-4k\frac{G(E)}{F(E)}-n\left(2+k\right)\right]
\nonumber \\ && \pm
\sqrt{\left\{4k\frac{G(E)}{F(E)}+n\left(2+k\right)-1\right\}^{2}-4n\left\{2\left(k+1\right) \left(n-1\right)-\left(5k+6\right)\right\}} ~~~~
\end{eqnarray}
Here $f_{1}(t)$ and $f_{2}(t)$ are arbitrary functions of $t$.

So,  the deformed metric (\ref{rainbow}) can be expressed as
\begin{eqnarray}\label{12}
ds^{2}&=&\frac{1}{F(E)^2}\left[-1+f_{1}(t)r^{\omega_{1}-1}+f_{2}(t)r^{\omega_{2}-1}\right.\nonumber \\ && \left. +\frac{2n\left(n-3\right)
\left(k+1\right)+3k\frac{G(E)}{F(E)}}{\left(1-\omega_{1}\right)\left(1-\omega_{2}\right)}-\frac{2n\lambda\left(\omega+k+2\right)}
{\left(2-\omega_{1}\right)\left(2-\omega_{2}\right)}r\right]dt^{2}
\nonumber \\ && + \frac{1}{F(E)G(E)}dtdr+\frac{1}{G(E)^2}r^{2}d\Omega_{2}^{2}
\end{eqnarray}
which is the Rainbow deformed  generalized Vaidya metric in
Galileon gravity.  {It may be noted that   this
solution represents a special class of solution of the general model. This is because the general solution was very complicated,
and so we made this assumption to simplify our analysis.
}

\section{Collapse Study}
In the previous section, we analyzed the Rainbow deformation of the   Vaidya metric in
Galileon gravity. In this section, we will analyze the gravitational collapse in this
theory. We can let $ds^{2}=0$ in Eq.  (\ref{rainbow}), and obtain the
  equation for outgoing radial null geodesics. It may be noted that
$d\Omega_{2}^{2}=0$, and
\begin{equation}\label{collapse2.24}
\frac{dt}{dr}=\frac{F(E)}{G(E)\left(1-\frac{m(t,r)}{r}\right)}.
\end{equation}
Thus, the central singularity exists at the  point   $r=0,~t=0$.
Now  we can study the behavior of the function  $X=\frac{t}{r}$ as it approaches this singularity at $r=0,~t=0$
along the radial null geodesic. Let us denote this    limiting value by
$X_{0}$, and so we can write
\begin{eqnarray}\label{collapse2.25}
\begin{array}{c}
X_{0}\\\\
{}
\end{array}
\begin{array}{c}
=lim~~ X \\
\begin{tiny}t\rightarrow 0\end{tiny}\\
\begin{tiny}r\rightarrow 0\end{tiny}
\end{array}
\begin{array}{c}
=lim~~ \frac{t}{r} \\
\begin{tiny}t\rightarrow 0\end{tiny}\\
\begin{tiny}r\rightarrow 0\end{tiny}
\end{array}
\begin{array}{c}
=lim~~ \frac{dt}{dr} \\
\begin{tiny}t\rightarrow 0\end{tiny}\\
\begin{tiny}r\rightarrow 0\end{tiny}
\end{array}
\begin{array}{c}
=lim~~ \frac{F(E)}{G(E)\left(1-\frac{m(t,r)}{r}\right)} \\
\begin{tiny}t\rightarrow 0\end{tiny}~~~~~~~~~~~~\\
\begin{tiny}r\rightarrow 0\end{tiny}~~~~~~~~~~~~
 {}
\end{array}
\end{eqnarray}
Now from   Eqs.  (\ref{10}) and (\ref{collapse2.25}), we obtain
\begin{eqnarray}
\frac{2}{X_{0}}=
\begin{array}llim\\
\begin{tiny}t\rightarrow 0\end{tiny}\\
\begin{tiny}r\rightarrow 0\end{tiny}
\end{array}\frac{2G(E)}{F(E)}\Bigg[1&-&f_{1}(t)r^{\omega_{1}-1}-f_{2}(t)r^{\omega_{2}-1}\nonumber \\&-&
\frac{2n\left(n-3\right)\left(k+1\right)+3k\frac{G(E)}{F(E)}}{\left(1-\omega_{1}\right)
\left(1-\omega_{2}\right)}+\frac{2n\lambda\left(\omega+k+2\right)}
{\left(2-\omega_{1}\right)\left(2-\omega_{2}\right)}\frac{r}{t}\Bigg]
\end{eqnarray}
Here  $f_{1}(t)=\delta t^{-(\omega_{1}-1)}$ ~and
~$f_{2}(t)=\epsilon t^{-(\omega_{2}-1)}$, where $\delta$ and $\epsilon$
are constants. Now  the equation for $X_{0}$ can be written as
\begin{equation}
\delta X_{0}^{2-\omega_{1}}+\epsilon
X_{0}^{2-\omega_{2}}-\left[1-\frac{2n\left(n-3\right)\left(k+1\right)+3k\frac{G(E)}{F(E)}}
{\left(1-\omega_{1}\right)\left(1-\omega_{2}\right)}\right]X_{0}+
2\left[1+\frac{n\lambda\left(\omega+k+2\right)}{\left(2-\omega_{1}\right)
\left(2-\omega_{2}\right)}\right]=0
\end{equation}
It may be noted that  the outgoing null geodesic exists for  $X_{0}>0$.
Thus,   a black hole will be formed when  none of  the solutions of this equation are
  positive. It is difficult to find analytic solutions for $X_{0}$, and so we will find
  numerical solutions for $X_{0}$. This will be done by assigning
specific  numerical values to the  constants associated with this model. We will also need to use a specific form
of the rainbow function for performing this numerical analysis. Thus, we will use
the    rainbow functions motivated from
  loop quantum gravity approach and   $\kappa$-Minkowski non-commutative spacetime \cite{Amelino1,
Amelino2},
\begin{equation}
F(E/E_{p})=1,~~~~~G(E/E_{p})=\sqrt{1-\eta\left(\frac{E}{E_{p}}\right)}
\end{equation}
 {In the above expressions, $E_{p}$ is the Planck energy and it is
given by $E_{p}=1/\sqrt{G}=1.221 \times 10^{19}$ GeV}. The behavior of the roots of this equation
can be obtained from  contour plots of
  $X_{0}$ vs $k$,   for fixed values of other
parameters. Thus, we will be able to understand the behavior of the
  collapse
at different cosmological eras. We can also understand the role played by other parameters in   the collapse  by
by adjusting the values of those parameters.
\begin{figure}
\includegraphics[height=1.6in]{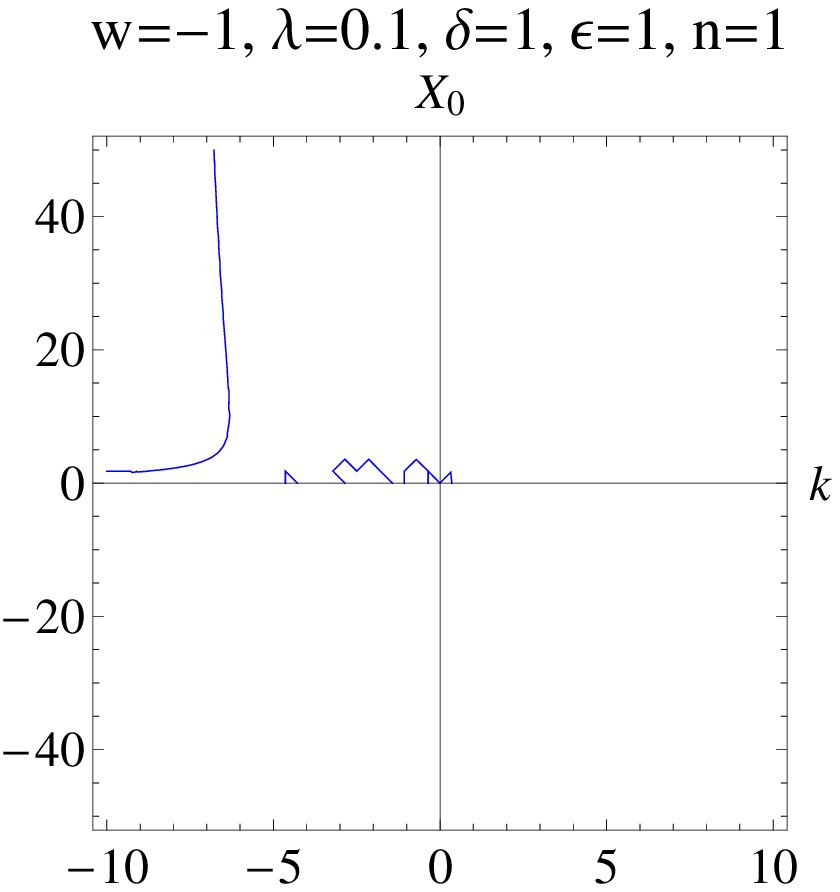}~~~~\includegraphics[height=1.6in]{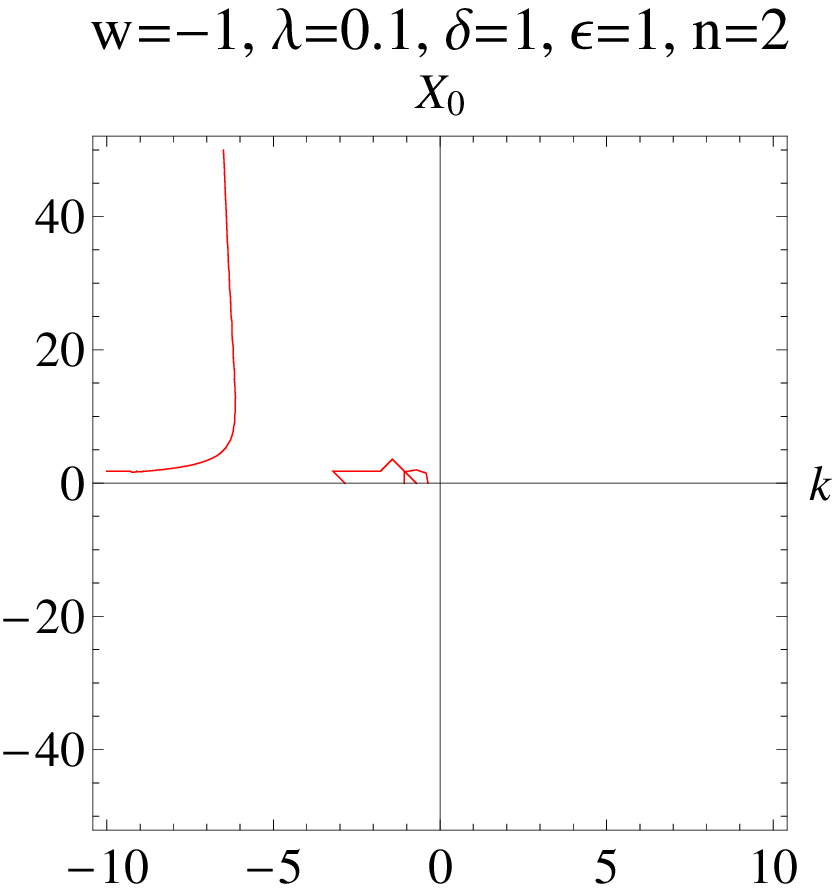}~~~~
\includegraphics[height=1.6in]{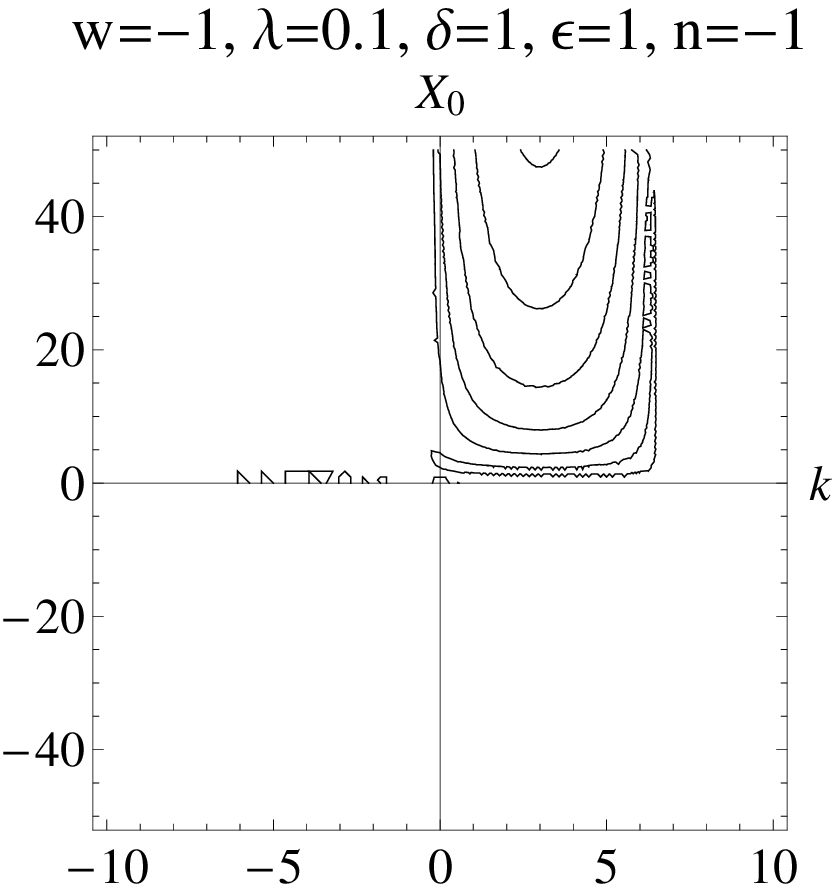}~~~~\includegraphics[height=1.6in]{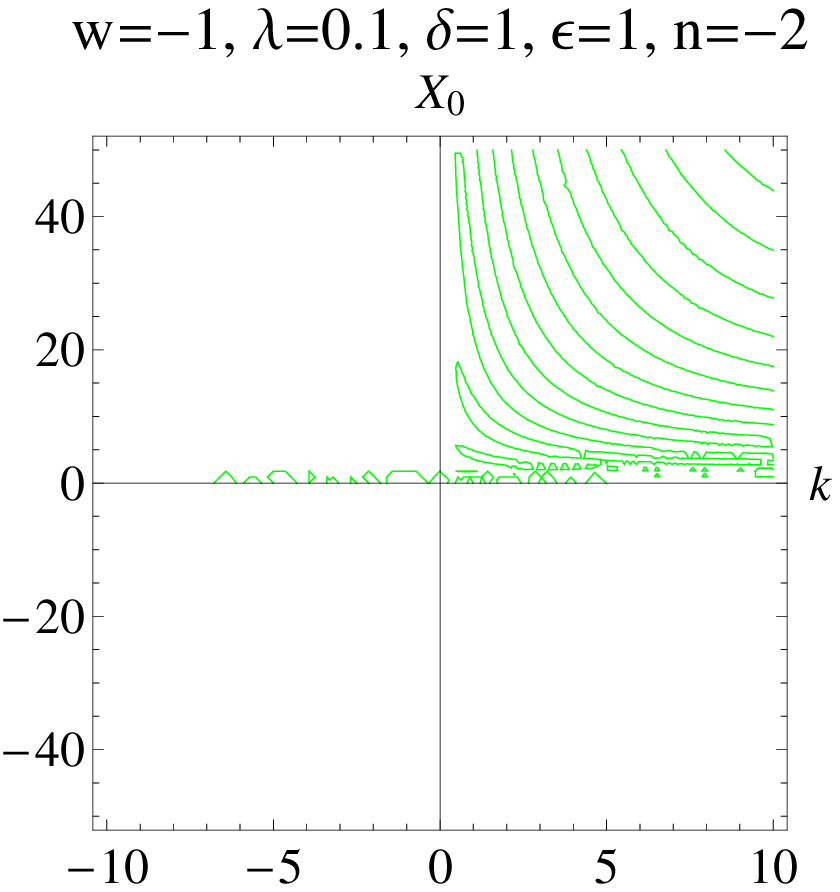}~~~~\\
\vspace{1mm}
~~~~~~~~~~~~~Fig.1~~~~~~~~~~~~~~~~~~~~~~~~~~~~~~~~~~~~~~~Fig.2~~~~~~~~~~~~~~~~~~~~~~~~~~~
~~~~~~~~~Fig.3~~~~~~~~~~~~~~~~~~~~~~~~~~~~~~~~~~~~~Fig.4~~~~~~~~~~~~~\\
\vspace{2mm} \textit{\textbf{Figs 1, 2, 3 and 4} show the
variation of $X_{0}$ with $k$ for different
values of $n$ and for $w=-1$}
\end{figure}

\begin{figure}
~~~~~~~~~~~~~~~~~~~~~~~~~~~~~~~~~~~~~~\includegraphics[height=1.6in]{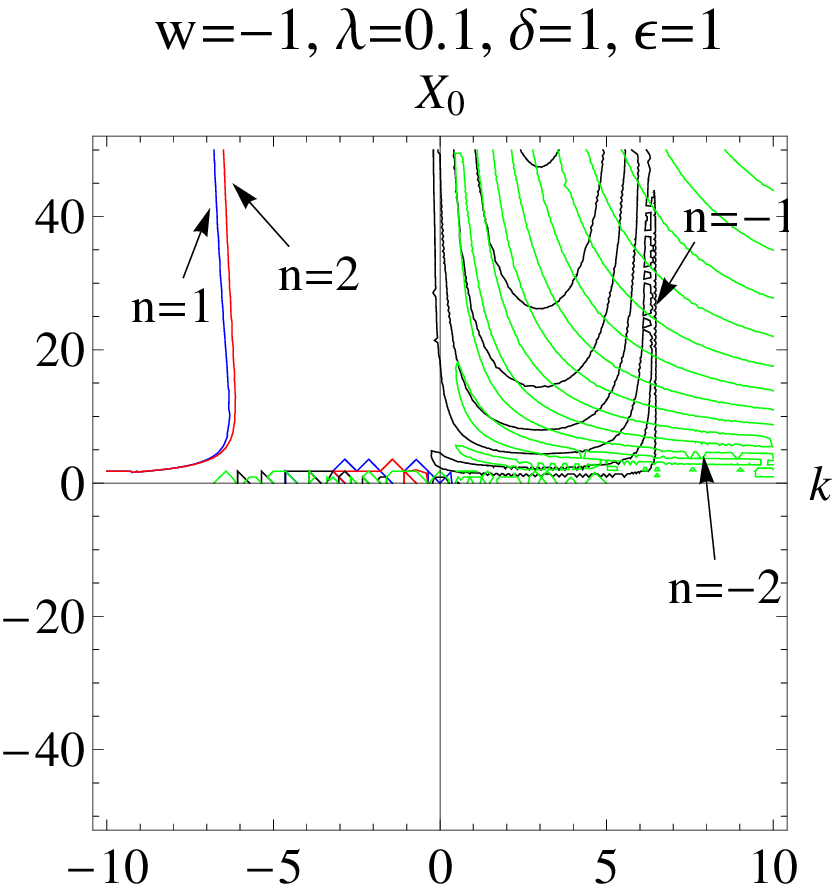}\\
\vspace{5mm}~~~~~~~~~~~~~~~~~~~~~~~~~~~~~~~~~~~~~~~~~~~~~~~~Fig.5~~~~\\
\vspace{2mm} ~~~~~~~~~~~~~~\textit{\textbf{Fig 5} shows the
combined effect of figs.1,2,3 and 4.}
\end{figure}

\begin{figure}
\includegraphics[height=1.6in]{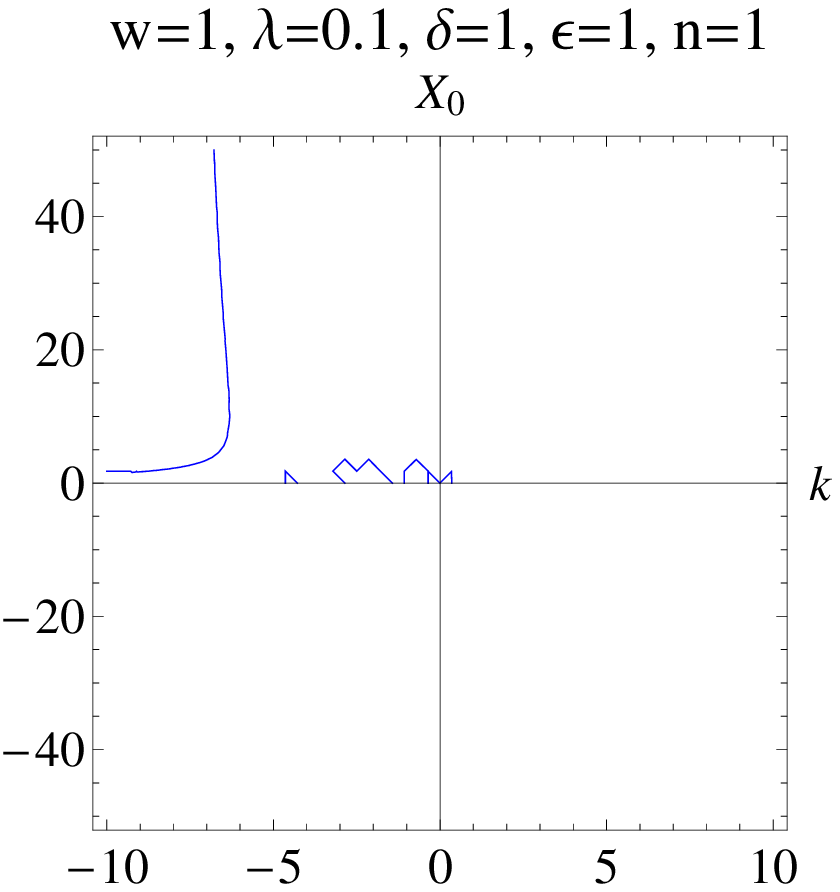}~~~~\includegraphics[height=1.6in]{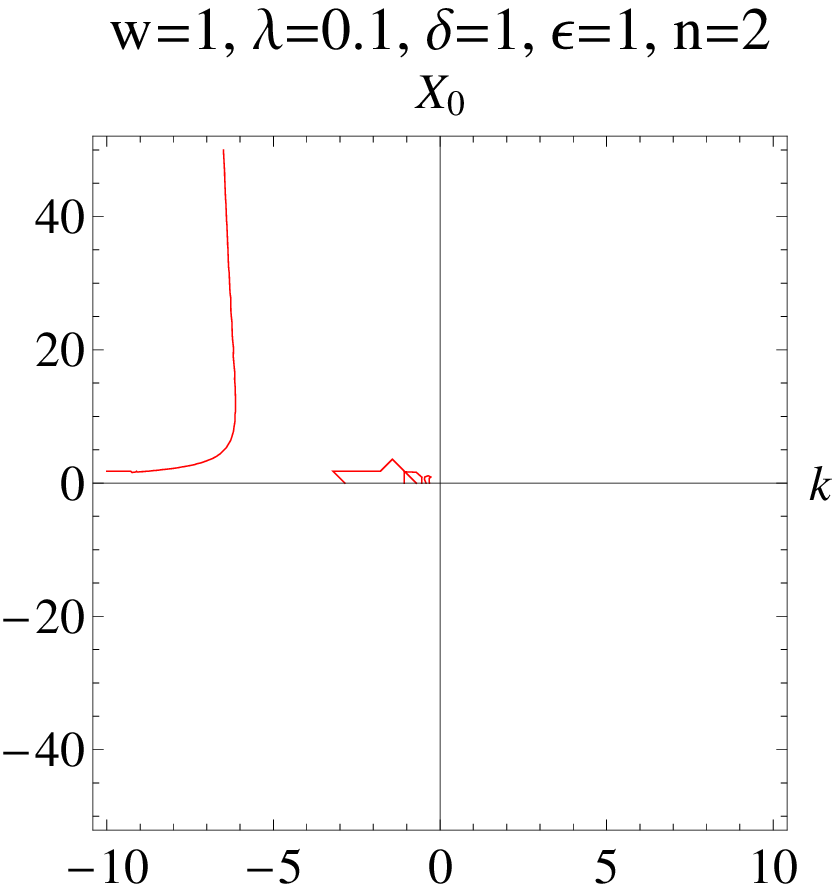}~~~~
\includegraphics[height=1.6in]{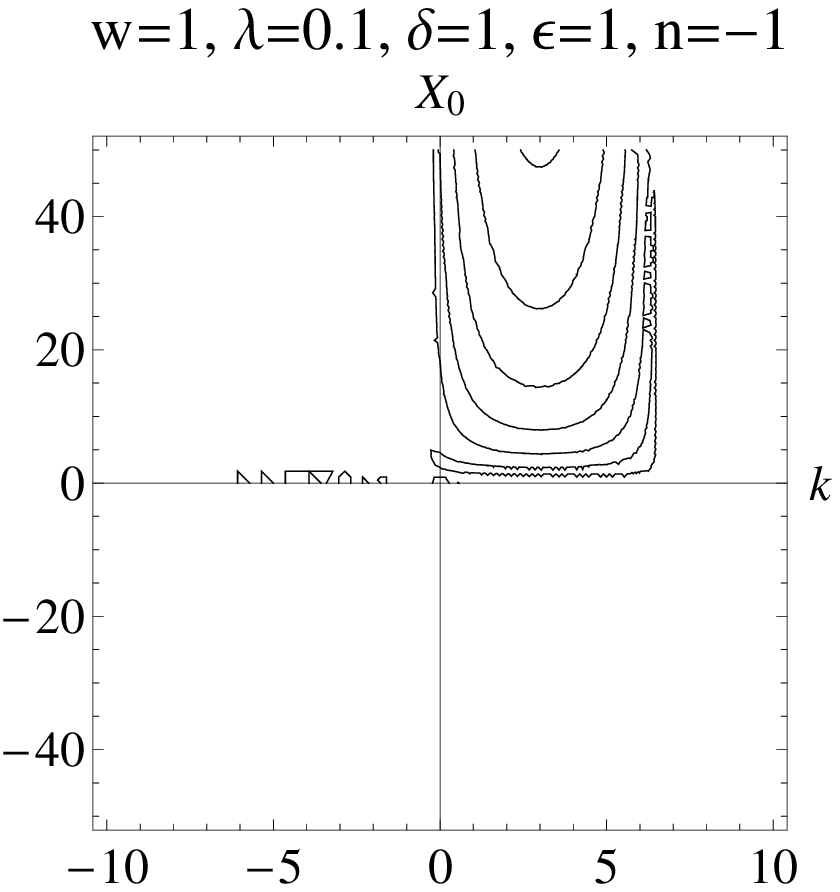}~~~~\includegraphics[height=1.6in]{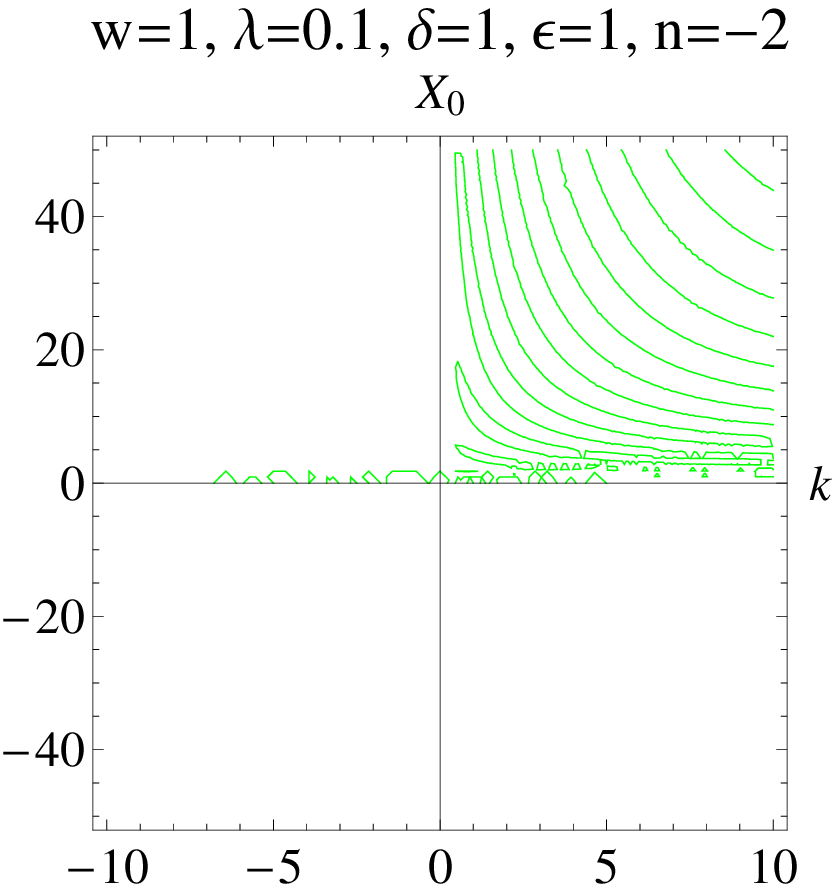}~~~~\\
\vspace{1mm}
~~~~~~~~~~~~~Fig.6~~~~~~~~~~~~~~~~~~~~~~~~~~~~~~~~~~~~~Fig.7~~~~~~~~~~~~~~~~~~~~~~~~~~~
~~~~Fig.8~~~~~~~~~~~~~~~~~~~~~~~~~~~~~~~~Fig.9~~~~~~~~~~~~~\\
\vspace{2mm} \textit{\textbf{Figs 6, 7, 8 and 9} show the
variation of $X_{0}$ with $k$ for different
values of $n$ and for $w=1$}
\end{figure}

\newpage
\begin{figure}
~~~~~~~~~~~~~~~~~~~~~~~~~~~~~~~~~~~~~~~~~\includegraphics[height=1.6in]{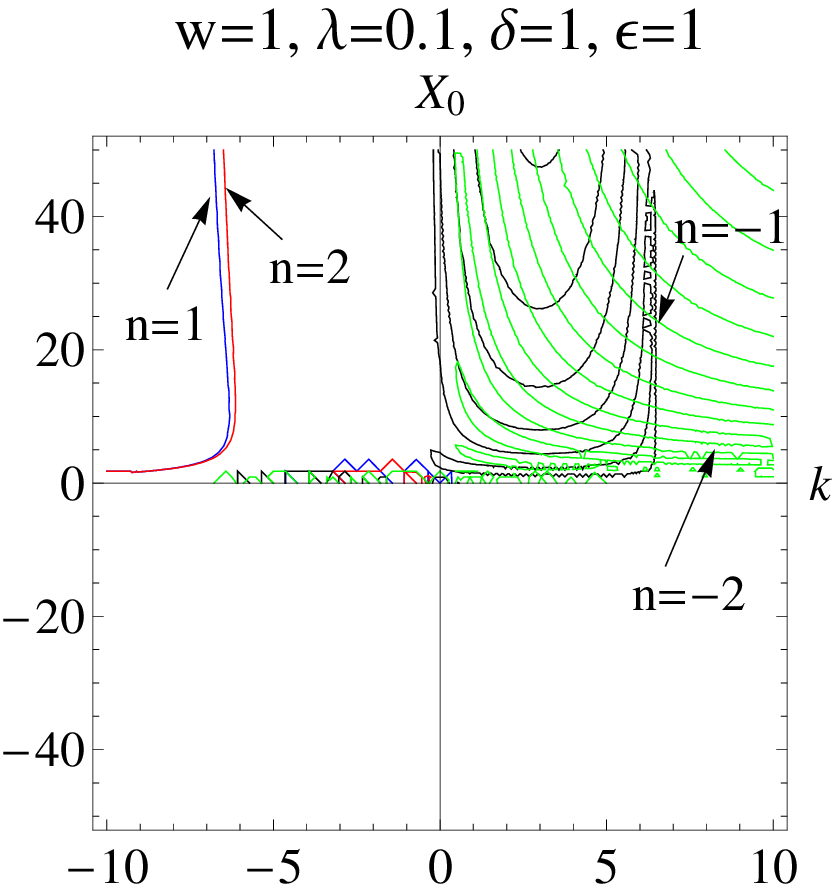}~~~~\\
\vspace{3mm}
~~~~~~~~~~~~~~~~~~~~~~~~~~~~~~~~~~~~~~~~~~~~~~~~~Fig.10~~~~~~~~~~~~~~~~~~~~~~~~~~~\\
\vspace{2mm} ~~~~~~~~~~~~~~~~~~~~\textit{\textbf{Fig 10} shows the
combined effect of figs.6,7,8 and 9.}
\end{figure}

\section{Discussions and Conclusions}
We have set the Galileon parameter $w=-1$ in
 figs. (1), (2), (3) and (4), to obtain the contour plots of $X_{0}$ vs $k$.
This has been done by using      different
values of $n$, and   fixed all   other parameters such as  $\delta$, $\lambda$,
and $\epsilon$. The positive solutions for $X_0$ can be obtained
in   different cosmological eras for all
  these cases.  Thus, in  figs. (1) and (2), it was observed that  for $n=1,2$, positive solutions
exist when  $k<-1$. So, for $n=1,2$,  positive solution exists in a phantom DE era
(late universe). However, it was also observed  in figs. (3) and (4), that for $n=-1,-2$, positive solutions exist when
  $k>0$. So, for $n=-1,-2$, positive solutions exist  in a radiation era
(early universe). We have also set the  Galileon parameter  $w=1$, and obtained similar results in
 figs. (6), (7), (8) and (9). It was observed that even the  range of $X_{0}$ was similar for $w=-1$
 and $w=1$.  Thus, the    collapsing system in
  rainbow deformed Galileon gravity does not depend on
the Galileon parameter $w$. A similar result was obtained form
the study of gravitational collapse
in the usual Galileon gravity \cite{v2}. It can be observed that
naked singularities are formed  in the late universe, and black holes are formed in the early universe,  for
positive values of $n$.   However, naked singularities are formed
in the early  universe, and black holes are formed in the late universe,  for
negative values of $n$.  Similar results are obtained from
 figs. (5) and (10)  for different
scenarios.  In this paper, we
first deformed Galileon gravity using rainbow functions. Then we analyzed the collapsing system in this Galileon gravity's rainbow.
It was observed that the
collapsing system does not depend on the  Galileon parameter $w$ in   rainbow deformed Galileon gravity.
It will be interesting
to analyze other systems using   a combination of gravity's rainbow with
Galileon gravity.

\section*{Acknowledgements}

The authors   acknowledge the anonymous referee for enlightening comments that helped to improve the quality of the manuscript.Ahmed Farag Ali is supported by STDF grant 13858 and by Benha University (www.bu.edu.eg).

\end{document}